\documentclass[11pt, A4]{article}
\usepackage[super, compress]{cite}
\usepackage{graphicx}
\usepackage[caption=false]{subfig}
\usepackage{amsmath}
\usepackage{amssymb}
\usepackage{enumitem}
\usepackage{url}    
\def\be{\begin{equation}}
\def\ee{\end{equation}}
\def\ba{\begin{eqnarray}}
\def\ea{\end{eqnarray}}

\newcommand\mca{\mathcal{A}}
\newcommand\mcb{\mathcal{B}}
\newcommand\mcc{\mathcal{C}}
\newcommand\mch{\mathcal{H}}

\newcommand\mba{\mathbf{a}}
\newcommand\mbb{\mathbf{b}}
\newcommand\bee{\begin{equation*}}
\newcommand\eee{\end{equation*}}
\def\be{\begin{equation}}
\def\ee{\end{equation}}
\def\ba{\begin{eqnarray}}
\def\ea{\end{eqnarray}}

\newcommand\dphi{\dot{\phi}}
\newcommand\eb{b_0 e^{-\alpha\phi}}

\begin{document}
\markboth{Staicova \& Stoilov}{Cosmology from multimeasure multifield model}

%
%

\title{Cosmology from multimeasure multifield model}

\author{ Denitsa Staicova$^1$ \and  Michail Stoilov$^2$}

\date{
Institute for Nuclear Research and Nuclear Energy \\Bulgarian Academy of Sciences, Sofia 1784, Tsarigradsko shosse 72, Bulgaria\\
$^1$ dstaicova@inrne.bas.bg\\
$^2$ mstoilov@inrne.bas.bg \\[2ex]%
    \today
}

\maketitle


\begin{abstract}
We consider the cosmological application of a (variant of) relatively newly proposed model \cite{1609.06915} unifying inflation, dark energy, dark matter, and the Higgs mechanism.
The model was originally defined using additional non-Riemannian measures but it can be reformulated into effective quintessential model unifying inflation, dark energy and dark matter.
Here we demonstrate numerically that it is capable of describing the entire Universe evolution in a seamless way, but this requires some revision of the model setup.
The main reason is that there is a strong effective friction in the model, a feature which has been neglected in the pioneer work.
This improves the model potential for proper description of the Universe evolution, because the friction ensures a finite time inflation with dynamically maintained low-value
slow-roll parameters in the realistic scenarios. In addition, the model predicts the existence of a constant scalar field in late Universe. 

\end{abstract}

\section{Introduction}
\label{intro}
According to the current understanding, we live in an expanding Universe. The rate of this expansion, however, is a hot topic for the community due to the so called $H_0$ tension -- i.e. the difference between the measurements of the Hubble constant locally (type IA supernova) and from the cosmic microwave background (the Planck mission). A recent study \cite{1903.07603} has shown that the $H_0$ tension is even more significant than expected -- the tension between the model-independent local measurements and the $\Lambda-CDM$ inferred Plank CMB measurements is about $4.4 \sigma$. Such a discrepancy calls for a revision of the default $\Lambda-CDM$ model and opens the door for theories of modified gravity. 

It is known that the Universe has passed trough different stages -- radiation-dominated era, matter-dominated era and current, dark-energy 
dominated era. Those epochs can be described well using the $\Lambda-CDM$ model as different components in the energy density of the Universe. Because of this, any new theory of gravity, needs to describe those stages just as well. Additionally, however, any new theory of gravity, has to be able to describe the initial inflation of the Universe -- the exponential expansion that is designed to solve the unresolved problems in cosmology --  the horizon problem, the flatness problem, the missing monopols problem and the large-structures formation problem \cite{cosmo0}). Numerous such theories have been proposed so far  -- chaotic inflation (one scalar field rolling in a potential), multi-field theories (more than one scalar fields), modified gravity (f(R), Brans-Dicke etc),  
for a review see \cite{Linde, Debono} , but they have seen their strongest limitation so far in the gravitational waves astronomy (for a review see \cite{1710.05901}). Among the few surviving theories is the {\em k-essence} type of theories of modified gravity, which do not modify the speed of the gravitational waves and thus are not in violation with the observational data. 

The considered in this article model is  in the framework of multi-measure gravitational models.
The basic idea is \cite{itb} that on the space-time manifold can be defined a metric-independent (non-Riemannian) volume form (or forms)  in terms of an auxiliary antisymmetric gauge field(s) of maximal rank.
Such forms can be used in the action integral to replace the usual integration measure  (or to participate together with it).
The approach was further developed in Refs.\cite{ref01,ref01_1, ref01_2, ref01_3} and successfully applied to define unified dark energy -- dark matter model \cite{dmde1, dmde2}, unified inflation -- dark energy model \cite{1407.6281, 1408.5344}, and a model unifying inflation, dark energy, dark matter, and Higgs mechanism \cite{1609.06915}.
 Some of these models are naturally connected to the $f(R)$ gravity and to the quintessential inflation proposed by  Peebles and Vilenkin \cite{Peebles}.
Effectively, the role of any such non-Riemannian measure is to produce very simple but powerful system of equations of motion for the quantity coupled to it in the Lagrangian.
As a result on-shell the corresponding quantity is a (dynamically generated) constant, determined by the problem boundary conditions.

In Ref. \cite{num_dark} we have studied the cosmological aspects of a model \cite{dmde2} with a single scalar field (called darkon) coupled to two independent volume-forms.
 The  model predicts  both the existence of dark energy (as a dynamically generated cosmological constant) and of dark matter (as a dust contribution to the energy-momentum tensor). 
The model is capable to fit the Supernova Type Ia data  still leaving a large freedom in its parameters. 
 Here we consider an ``upgrade'' of the darkon model aiming to include initial inflation.
The new model features two scalar fields (darkon and inflaton) and has been introduced in Ref.\cite{1609.06915}.
In this model, the inflaton moves over an effective step-like potential (which plays the role of cosmological constant) with two infinite plateaus connected with a steep slope. 

 It is advocated in Refs. \cite{1408.5344, 1609.06915} that the left, higher plateau corresponds to the inflationary Universe, while the right, lower plateau -- to the current expanding Universe. 
Our results show that for some particular values of the parameters we can obtain all epochs in the evolution of the Universe, including a graceful exit to current exponential expansion.
However, the results demonstrate the existence and importance of a friction term in the inflaton equation of motion.
As a consequences it is impossible to start the evolution from the left plateau and to reach the right plateau regardless of the initial inflaton velocity.  The early inflation, instead, is generated on the slope of the effective potential after a brief ultra-relativistic period due to the initial singularity. While the large parameter-space has not been fully explored, the cases presented here are able to demonstrate the qualitatively different behaviors of the numerical solution.

\section{The model}
We work with a simplified version (without Higgs and gauge fields) of the model proposed in Ref. \cite{1609.06915}.
The model itself is based on earlier results on dark matter -- dark energy unified model \cite{dmde1, dmde2} and inflation -- dark energy unified model  \cite{1407.6281, 1408.5344}. 
We have two scalar fields --- the darkon $u$ and inflaton $\phi$ and the action of the  is the following:
\be
S=  S_{darkon}+S_{inflaton}.
\label{totA}
\ee
Here (in units $G_N=1/16 \pi,\; c=1$):
\ba
  S_{darkon}&=&\int{d^4x\;(\sqrt{-g}+\Phi(\mcc))L^{(0)}},
\label{darA}\\
S_{inflaton}&=&\int d^4x\; \Phi(\mca)(R+L^{(1)})+\int d^4x\;\Phi(\mcb)\left(L^{(2)}+\frac{\Phi(\mch)}{\sqrt{-g}}\right).
\label{infA}
\ea
We use the notations
\ba 
L^{(0)}&=&Y -W(u),\label{dla}\\
L^{(1)}&=&X- V(\phi),\label{lla1}\\
L^{(2)}&=&\eb X + U(\phi),\label{lla2}
\ea
where $ Y= -\frac{1}{2}g^{\mu\nu}\partial_\mu u\partial_\nu u$ and $X=-\frac{1}{2}g^{\mu\nu}\partial_\mu\phi\partial_\nu \phi$ are the darkon and inflaton kinetic terms respectively, and 
$V(\phi)=f_1 e^{-\alpha \phi},\;U(\phi)=f_2 e^{-2\alpha \phi}$ and $W(u)$ are corresponding potentials. 
This particular form of the inflaton potentials ensures a global Weyl invariance of $S_{inflaton}$.
On the other hand it turns out that the equations of motion do not depend on the form of the darkon potential $W(u)$ so we do not specify it.
Note that we use $R=g^{\mu\nu}R_{\mu\nu}(\Gamma)$ written in Palatini formalism where  $\Gamma^\lambda_{\mu\nu}$ are treated as {\it a priori} independent variables.

\noindent The quantity $\Phi(Z)$, where $Z=\mca,   \mcb, \mcc, \mch$, which appears in Eqs. \ref{darA}, \ref{infA} is an additional non-Riemannian generally covariant integration  measure defined with the help of an auxiliary 3-index antisymmetric tensor gauge field $Z_{\nu\kappa\lambda}$: 
\be
\Phi(Z)=\frac{1}{3\!}\epsilon^{\mu\nu\kappa\lambda}\partial_\mu Z_{\nu\kappa\lambda}.
  \label{TMG}
\ee
Thus the model depends on $4$ parameters --- dimensionless $b_0$ and dimensional $f_1, f_2$ and $\alpha$.
The action in Eq.\ref{infA} is exactly the same as that in Refs. \cite{1407.6281, 1408.5344}, while the action in Eq. \ref{darA} differs from that in   \cite{dmde2} which is 
$S=\int{d^4x\;\sqrt{-g}(R-\epsilon R^2)} + S_{darkon}$
because the first terms in the latter already present in $S_{inflaton}$  in a hidden form.

\subsection{The model equations of motion}

Our primary objective is to use the equations of motion for the auxiliary fields to eliminate them and to obtain an effective Lagrangian written down entirely in terms of metric and scalar fields.
Probably the only unfamiliar equations of motion are those obtained by variation of the action with respect to the 3-forms  $Z=\mca, \mcb, \mcc, \mch$. 
The easiest way to get them is to note that any such $Z$  has only 4 independent components, namely 
$\hat{z}^\mu=\epsilon^{\mu\nu\kappa\lambda} Z_{\nu\kappa\lambda}/3$.
Written in terms of $\hat{z}^\mu$ the integration measure is $\Phi(Z)=\partial_\mu\hat{z}^\mu$ and now it is trivial to obtain the variation of the action $S'=\int{d^4x\;\Phi(Z)L'}$ (this is the most general form of the action with the non-Riemannian measures we use) with respect to independent components of $Z$:
\be 
0=\frac{\delta S'}{\delta \hat{z}^\mu}=
\frac{\delta \int{d^4x\;(\partial_\nu \hat{z}^\nu)L'}}{\delta \hat{z}^\mu}\;\; 
\Rightarrow \partial_\mu L'=0\;\; \forall \mu.
\ee
Thus, the solution of the $Z$-equation of motion is $L'=\mathrm{constant}$. 
We have four auxiliary 3-forms in Eq.\ref{totA}, so we obtain four dynamically generated integration constants:
\ba
 L^{(0)}&=&-2M_0, \label{co1}\\
 R+L^{(1)}&=&M_1, \label{co2}\\
 L^{(2)}+\frac{\Phi(\mch)}{\sqrt{-g}}&=&-M_2,\label{co3}\\
 \frac{\Phi(\mcb)}{\sqrt{-g}}&=&\chi_2
 \label{const} 
\ea
Note that in the above equations only $\chi_2$ is dimensionless, while $M_0, M_1$ and $M_2$ are with dimension $\mathrm{mass}^4$, the same as $f_1$ and $f_2$.

The full-fledged derivation of the equation of motion for $\Gamma^{\alpha}_{\beta_\gamma}$ and its solution can be found in Ref.\cite{ref01} .
The result is that $\Gamma$ is the Levi--Civita connection 
\be
\Gamma^\mu_{\lambda\kappa}=\frac{1}{2}\tilde{g}^{\mu\nu}(\partial_\nu \tilde{g}_{\lambda\kappa}-\partial_\lambda \tilde{g}_{\nu\kappa}-\partial_\kappa \tilde{g}_{\nu\lambda})
\label{lcc}
\ee
 for the Weyl rescaled metric 
$\tilde{g}_{\mu\nu}$:
\be 
\tilde{g}_{\mu\nu}=\chi_1 g_{\mu\nu},\label{amet}
\ee where
\be
\chi_1=\frac{\Phi(\mca)}{\sqrt{-g}}.\label{ch1}
\ee
There is a simple way in two steps not to derive but to anticipate the result given in Eq. \ref{lcc}.
First, note that from Eq.\ref{amet} we get $\Phi(\mca)g^{\mu\nu}=\sqrt{-\tilde{g}}\tilde{g}^{\mu\nu}$,
so the term $\int d^4x\; \Phi(\mca) R=\int d^4x\; \Phi(\mca) g^{\mu\nu}R_{\mu\nu}(\Gamma)$ in Eq.\ref{infA} looks exactly as the Einstein--Hilbert action in Palatini formalism with metric $\tilde{g}_{\mu\nu}$.
Second, under some general assumptions, the solution of Palatini formalism is the corresponding Levi-Civita connection Eq.\ref{lcc} Q.E.D.  

Eq.\ref{lcc} signals that the effective Lagrangian we are looking for has to be written in terms of $\tilde{g}_{\mu\nu}$.
More precisely, we are looking for $L^{(eff)}$ and effective action 
\be S^{(eff)}=\int{d^4x \;\sqrt{-\tilde{g}}(\tilde{R}+L^{(eff)})}\label{sef}\ee
such that the corresponding Einstein equation
\be 
\tilde{R}_{\mu\nu}-\frac{1}{2}\tilde{g}_{\mu\nu}\tilde{R}= \frac{1}{2} T^{(eff)}_{\mu\nu}
\label{ei11}
\ee
 is equivalent to the equation of motion obtained by varying action Eq. \ref{totA} with respect to  $g^{\mu\nu}$, which is:
\ba 
0&=& \Phi(\mca) R_{\mu\nu} + \frac{\Phi(\mca)}{2}(L^{(1)}g_{\mu\nu} -T^{(1)}_{\mu\nu})+
\frac{\Phi(\mcb)}{2}(L^{(2)}g_{\mu\nu} -T^{(2)}_{\mu\nu})+
\frac{\Phi(\mcb)\Phi(\mch)}{2 \sqrt{-g}}g_{\mu\nu}\nonumber\\
&&-\frac{\sqrt{-g}}{2}L^{(0)} g_{\mu\nu}+\frac{\sqrt{-g}}{2}(L^{(0)} g_{\mu\nu}-T^{(0)}_{\mu\nu})+
\frac{\Phi(\mcc)}{2}(L^{(0)} g_{\mu\nu}-T^{(0)}_{\mu\nu}).
\label{EE0}
\ea
Here $\tilde{R}_{\mu\nu}$ and $\tilde{R}=\tilde{g}^{\mu\nu}\tilde{R}_{\mu\nu}$ are the Ricci tensor and scalar respectively for metric $\tilde{g}_{\mu\nu}$, 
$T^{(eff)}_{\mu\nu}\equiv \tilde{g}_{\mu\nu}L^{(eff)} - 2 \partial L^{(eff)}/\partial \tilde{g}^{\mu\nu}$, and
$T^{(i)}_{\mu\nu}\equiv g_{\mu\nu}L^{(i)} - 2 \partial L^{(i)}/\partial g^{\mu\nu},\;\;
i=0,1,2$.

Using Eqs.\ref{co2}, \ref{co3}, and the notation $\chi_3=\Phi(\mca)/\sqrt{-g}$, we put Eq.\ref{EE0} in the following form:
\be
R_{\mu\nu}-\frac{1}{2}R g_{\mu \nu}=
\frac{1}{2}T^{(1)}_{\mu\nu} +
\frac{\chi_2}{2\chi_1}T^{(2)}_{\mu\nu}
+\frac{1+\chi_3}{2\chi_1}T^{(0)}_{\mu\nu}
+\left(-\frac{M_1}{2}+\frac{\chi_2 M_2}{2\chi_1}+ \frac{\chi_3 M_0}{\chi_1}\right)g_{\mu\nu}.
\label{EE}
\ee
We need two more relations, both following from Eq.\ref{ch1}, in order to obtain $L^{(eff)}$.
First, because  by definition $\tilde{R}_{\mu\nu}=R_{\mu\nu}$, so, $\tilde{R}=R/\chi_1$ and therefore $\tilde{R}_{\mu\nu}-\frac{1}{2}\tilde{g}_{\mu\nu}\tilde{R}=R_{\mu\nu}-\frac{1}{2}R g_{\mu \nu}$.
Second, $(\tilde{g}_{\mu\nu} - 2 \partial /\partial \tilde{g}^{\mu\nu})L^{(eff)}= 
\chi_1(g_{\mu\nu} - 2 \partial /\partial g^{\mu\nu})L^{(eff)}$. Then, it is straightforward  to check that $L^{(eff)}$ written in terms of $g_{\mu\nu}$ is;
\ba 
\chi_1 L^{(eff)} &=& L^{(1)}-M_1 +\frac{\chi_2}{\chi_1}(L^{(2)}+M_2)+
\frac{1+\chi_3}{\chi_1}(L^{(0)}+2 M_0)-\frac{2 M_0}{\chi_1}\nonumber\\
&=& L^{(1)}-M_1 +\frac{\chi_2}{\chi_1}(L^{(2)}+M_2)-\frac{2 M_0}{\chi_1},
\ea
where we have used Eq.\ref{co1} to obtain the final form.
In order to write $L^{(eff)}$ as a function depending on $\tilde{g}_{\mu\nu}$ we introduce the notations
\ba 
u\rightarrow\tilde{u}&:&  \frac{\partial\tilde{u}}{\partial u}=(W-2M_0)^{-\frac{1}{2}}
\label{ut}\\
\tilde{Y}&=&-\frac{1}{2}\tilde{g}^{\mu\nu}\partial_\mu\tilde{u} \partial_\nu\tilde{u} 
\label{yt}\\
\tilde{X}&=&-\frac{1}{2}\tilde{g}^{\mu\nu}\partial_\mu\phi \partial_\nu\phi 
\label{xt}
\ea
The change of variable Eq. \ref{ut} has been used in Refs.\cite{dmde1, dmde2} to demonstrate that the darkon model represents relativistic fluid.
An important consequence of  Eqs. \ref{ch1}, \ref{ut}, \ref{yt} is that 
\be \tilde{Y}=1/\chi_1.\label{ych}\ee
Taking into account all this, we obtain
\be
L^{(eff)}=\tilde{X}-\tilde{Y}(V+M_1-\chi_2\eb\tilde{X})+\tilde{Y}^2(\chi_2(U+M_2)-2M_0).
\label{l_new}
\ee
This Lagrangian is non-linear with respect to both scalar fields kinetic terms and thus can be classified to be of generalized k-essence type. 

The effective Lagrangian possesses an obvious global symmetry $\tilde{u}\rightarrow\tilde{u}+w$ which produces the following current conservation
\be
\partial_\mu\left(\sqrt{-\tilde{g}}\tilde{g}^{\mu\nu}\partial_\nu\tilde{u}\frac{\partial L^{(eff)}}{\partial \tilde{Y}}\right)=0.
\label{cons_curr}
\ee
which coincides with the equation of motion for $\tilde{u}$.

 The variation with respect to the scalar field $\phi$ is easier done in a specific metric and we postpone it to subsection 2.3.

\subsection{The static case}
Taking the trace of Eq.\ref{EE0} we get:
\be 
\frac{1}{\chi_1}=\frac{2 (L^{(1)} +M_1 -T^{(1)}/2)}
{\chi_2(T^{(2)} +4 M_2 )+ (1 +\chi_3)T^{(0)} + 8 \chi_3 M_0}
\ee
which in the case when the fields $u$ and $\phi$ are static and taking into account Eq.\ref{ych} gives:
\be 
\tilde{Y}^{static}=\frac{V+M_1}{2\chi_2(U+M_2)-4M_0}.
\ee
Substituting the above equation into Eq.\ref{l_new} we see that the effective Lagrangian determines the following effective potential:
\be
U_{eff}=\frac{1}{4}\frac{(V+M_1)^2}{\chi_2(U+M_2)\!-\!2M_0}.
\label{ueff}
\ee
This potential for certain parameters looks like two infinite plateaus with asymptotic values $U_{-}=U_{eff}\vert_{\phi\rightarrow -\infty}=f_1^2/4\chi_2 f_2$ and  $U_{+}=U_{eff}\vert_{\phi\rightarrow +\infty}=M_1^2/(4\chi_2 M_2-8M_0)$ and a steep slope connecting them.
This means that the model naturally determines two different cosmological constants when 
$\phi\to \pm\infty$.
However, there is a catch in the usage of the effective potential ---
the effective potential does not bring the kinetic energy in standard form. 
We write more on this topic at the end of next subsection.

\subsection{The model in the Friedmann -- Lema\^itre -- Robertson -- Walker (FLRW) metric}

In the metric $\tilde{g}_{\mu\nu}=\mathrm{diag}\{-1, a(t)^2, a(t)^2, a(t)^2\}$ and with fields depending only on the time coordinate $(t)$, Eq.\ref{sef} takes the form 
\ba
S^{(eff)}\vert_{FLRW}
&=&\int dt \;a(t)^3\Big(-6\;\frac{\dot{a}(t)^2}{a(t)^2}
+\frac{\dot{\phi}^2}{2}
-\frac{v^2}{2}\left(V+M_1-\chi_2\eb\dot{\phi}^2/2\right)\Big.\nonumber\\
&& \Big. +\frac{v^4}{4}\left(\chi_2(U+M_2)-2M_0\right)\Big).
\ea
Here we use the notation $v=d\tilde{u}/dt$) and have performed the integration by parts when writing the term 
$\int{d^4x \;\sqrt{-\tilde{g}}\tilde{R}}$.

The equations of motion and their solutions are relatively simple in the FLRW metric.
From Eq. \ref{cons_curr} we obtain the following cubic equation with respect to $v$: 
\ba
v\left( V+M_1-\frac{1}{2}\chi_2 \eb \dot{\phi}^2\right)+v^3 (\chi_2(U(\phi)+M_2)-2M_0)=\frac{p_u}{a(t)^3} \label{cubic}
\ea 
where $p_u$ is an integration constant. 
The  $\phi$-equation of motion is: 
\be
\frac{d}{dt}\left( a(t)^3\dphi(1+\frac{ v^2}{2}\chi_2\eb) \right)+
a(t)^3 (\alpha\frac{v^2}{4}\chi_2\eb \dphi^2+\frac{v^2}{2}V_\phi-\chi_2\frac{v^4}{4} U_\phi)=0
\label{fe}
\ee
where $U_\phi=\frac{\partial U}{\partial \phi}$,
 $V_\phi=\frac{\partial V}{\partial \phi}$. 

 The  Friedmann equations for a perfect fluid with density $\rho$ and pressure $p$  are: 
\ba
&\frac{\dot{a}(t)^2}{a(t)^2}&=\frac{\rho}{6}\label{ffe}\\
&\frac{\ddot{a}(t)}{a(t)}&=-\frac{3p+\rho}{12}\label{sfe}
\ea
where
\ba
&p&=\frac{1}{2}\dphi^2 (1+\frac{1}{4}\chi_2\eb v^2)-\frac{1}{4}v^2(V+M_1)+
\frac{p_u v}{4 a(t)^3}\\
&\rho&=\frac{1}{2}\dphi^2 (1+\frac{3}{4}\chi_2 \eb v^2)+\frac{v^2}{4} (V+M_1)+
\frac{3 p_u v}{4a(t)^3}.
\label{rho_p}
\ea
We will use only the first Friedmann equation (Eq. \ref{ffe}), because the second one (Eq. \ref{sfe}) is a consequence of Eqs. \ref{cubic}, \ref{fe} and \ref{ffe}.

Note that the model in FLRW metric has a strong friction term. 
To see this, one can decouple Eqs. \ref{cubic}--\ref{ffe} assuming that the system is in exponentially growing regime $a(t)=e^{H t}$.
In this case Eq. \ref{fe} can be put in the form:
\be\ddot{\phi}(t)+ 3\dot{\phi}(t)H=0\label{fric}\ee
 with a general solution 
 $$\phi(t) = C_1+C_2 e^{-3H t}.$$
We see that the friction is entirely determined by the Hubble parameter $H$.
As a result one cannot start from the left plateau  and reach the right plateau  of the effective potential a feature we also observe numerically. 
The friction ensures that in the far future the kinetic energy of the $\phi$ field vanishes and  we have  a dynamically generated cosmological constant 
\be
\Lambda_{eff}=\frac{U_+}{2}=\frac{M_1^2}{8(\chi_2 M_2-2M_0)}\label{coscon}.
\ee
As it is seen form Eq. \ref{fric} the existence of friction is independent of the model parameters and it is a feature of FLRW metric. 

It has been already mentioned the model possesses an effective potential depending only on the inflaton field  (see Eq. \ref{ueff}).
Unfortunately the effective potential
does not bring the kinetic energy in standard form.  
For example in the limit of slow roll approximation (neglecting the terms $\sim\;\dot{\phi}^2,\;\; \dot{\phi}^3,\;\; \dot{\phi}^4$) the inflaton equation has the form: 
\be 
(1+A)\ddot{\phi}+3H(1+A)\dot{\phi}+U_{eff}'=0,
\label{eq_infl}
\ee 
where $A=v(t)^2 b\,\chi_2 e^{-\alpha \phi}/2$ and $H=\dot{a}(t)/a(t)$. One can see that putting the kinetic term in standard form (i.e., dividing Eq. \ref{eq_infl} by $(1+A)$)  introduces a singularity at $a=0$ in the effective potential term (due to $v\sim p_u/a^3$). This changes significantly the behavior of the system in the region $a\sim 0$ as seen in Ref.\cite{1808.08890} . Because of this problem, we will not use the effective potential, but will stick in our calculations to the system Eqs. \ref{cubic}--\ref{ffe}.

\section{Numerical results}
Putting all things together, the algebraic--differential system we have to solve is:
\ba
v^3+3\mba v+2\mbb&=&0 \label{sys1}\\
\dot{a}(t)-\sqrt{\frac{\rho}{6}}a(t)&=&0 \label{sys2}\\
\frac{d}{dt}\left( a(t)^3\dphi(1+\frac{\chi_2}{2}\eb v^2) \right)+\hspace{25mm}&&\nonumber\\
a(t)^3 (\alpha\frac{\dphi^2}{4}\chi_2\eb v^2+\frac{1}{2}V_\phi v^2-\chi_2 U_\phi\frac{v^4}{4})&=&0\label{sys3}
\ea
where  the parameters of the cubic equation are:

$\mba=-(V+M_1-\chi_2 \eb \dot{\phi}^2/2)/3(\chi_2(U+M_2)-2M_0)$,  
$\mbb=-p_u/2a(t)^3(\chi_2(U+M_2)-2M_0)$.
 
Note that Eq.(\ref{sys1}) always has at least one real root but there is no global continuous definition of it on $\{\mba, \mbb\}$ plane. The root we use in this article is:
$$v=(-\mbb+\sqrt{\mba^3+\mbb^2})^{\frac{1}{3}}-\frac{\mba}{(-\mbb+\sqrt{\mba^3+\mbb^2})^{\frac{1}{3}}},$$ 
and a special care is taken that this root is {\em real} in the domain where we use it.

The system Eqs.(\ref{sys1}--\ref{sys3}) is of second order with respect to $\phi$ and we shall integrate it numerically\footnote{For the numerical solution of Eqs.(\ref{sys2}--\ref{sys3}) we use Fehlberg fourth-fifth order Runge-Kutta method with degree four interpolation implemented in Maple.}. 
In order to put it on a computer we have to fix our units.
We use $c=1$, $G_N=1/16\pi$ and $t_u=1$, where $c$ is the speed of light, $G_N$ is Newton constant and $t_u$ is the present day age of Universe ($t_u=13.8 10^9\; y$). 
Thus, our mass unit is  $1.62 \times 10^{59}M_{Pl}$ and the cosmological constant ($\Lambda=1.9 10^{-35}s^{-2}$)  is $\Lambda\approx 3.6$. The modern-day Hubble constant in these units equals to $H_0\approx 0.95$.  Also, in these units,  the three periods of the evolution of the universe are : exit from early inflation: $t_{\mathrm{SD}} \sim 10^{-50}$ and start of present day acceleration  $t_{\mathrm{AE}} \sim 0.71$.

Altogether in our system  we have 4 free parameters, namely 
$\{\alpha, b_0, f_1, f_2\}$, 
5  integration constants $\{M_0, M_1, M_2, \chi_2, p_u\}$ and three initial conditions  $\{a(0), \phi(0), \dot \phi (0)\}$
\footnote{Note that the differential equation Eq. \ref{sys2} has a singular point in ${a}(0) = 0 $ due to the term $\sim 1/a(t)^3$ in $\rho$. 
So, a natural replacement of the initial condition  $a(0)=0$ is the normalization condition $a(1)=1$, where $t=1$ corresponds to the current moment.}.
We take some  steps to constrain the used parameters:

--  In order to ensure that we are working with a real root of Eq.(\ref{sys1}) , we choose  $ M_0<0$ (i.e. $\mbb<0$).
Moreover, we use $M_0\sim -0.01$.
 We are free to fix $M_0$ in this way, because as we have shown in Ref. \cite{num_dark} in the asymptotic darkon-dominated late-time inflation universe, we have a nice fit of Supernovae Ia data for any $M_0$.

-- In this article we work with $\dot{\phi}(0)=0$. In the general case, due to the strong friction, the system is not very sensitive to changes in  $\dot{\phi}(0)$.

--  We use $b_0>0$ to avoid problems with $\rho$, because $b_0<0$ can lead to negative kinetic term in $\rho$.

-- It is assumed in Ref.\cite{1408.5344} that the left plateau of the effective potential corresponds to the pre-inflationary Universe (Planck times) and the right plateau to the current and future accelerated expansion. Since we do not observe a solution for the inflaton connecting the two plateaus, the only restriction we apply to the model is that the left plateau is higher than the right one, i.e.
\be
\frac{f_1^2}{\chi_2 f_2}>>\frac{M_1^2}{\chi_2 M_2-2M_0}.
\label{res1}
\ee
Moreover, the estimates in Ref.\cite{1408.5344} for the values of the ratio $f_1^2/(\chi_2 f_2)$ have to be considered as lower limit.

We have presented already \cite{1801.07133}  some results from the numerical investigation of the model. 
Here we focus our attention on some unexpected features of the model which may be useful for its further theoretical improvement.
We observe that the solution of the system Eqs. \ref{sys1}--\ref{sys3} critically depends on the initial value of the inflaton field.
If $\phi(0)$ is out of some very specific parameter dependent interval
(which is always within the slope of effective potential), 
we obtain two-stage evolution of the Universe beginning with short deceleration followed by infinite acceleration.
When  $\phi(0)$ is in the interval we have four stage evolution which we refer to as``physically realistic''.
We have a short first deceleration epoch (FD), early inflation (EI), second deceleration (SD) which we interpret as radiation and matter dominated epochs together and finally we have infinite accelerating expansion (AE). 
There are also some very specific "pathological" parameter configurations with only deceleration.
The physically realistic parameter configurations are just a small part of the  parameter space, very sensitive to fine-tuning. Thus, a significant part of the numerical work is to find the points in the parameter space for which we observe the physically realistic behavior. This means that we have to find solutions for which the second derivative of the scale factor crosses the t-axis in three points ($t_{EI}, t_{SD}$ and $t_{AE}$), turning the problem into root-finding problem. This can be done by plotting the function or if we require higher precision, it can be done by root-finding algorithms such as the 1-d Muller algorithm applied to the polynomial approximation of $\ddot{a}(t)$. 

\begin{figure}[!h]\centering{
      \subfloat{{\includegraphics[width=4cm]{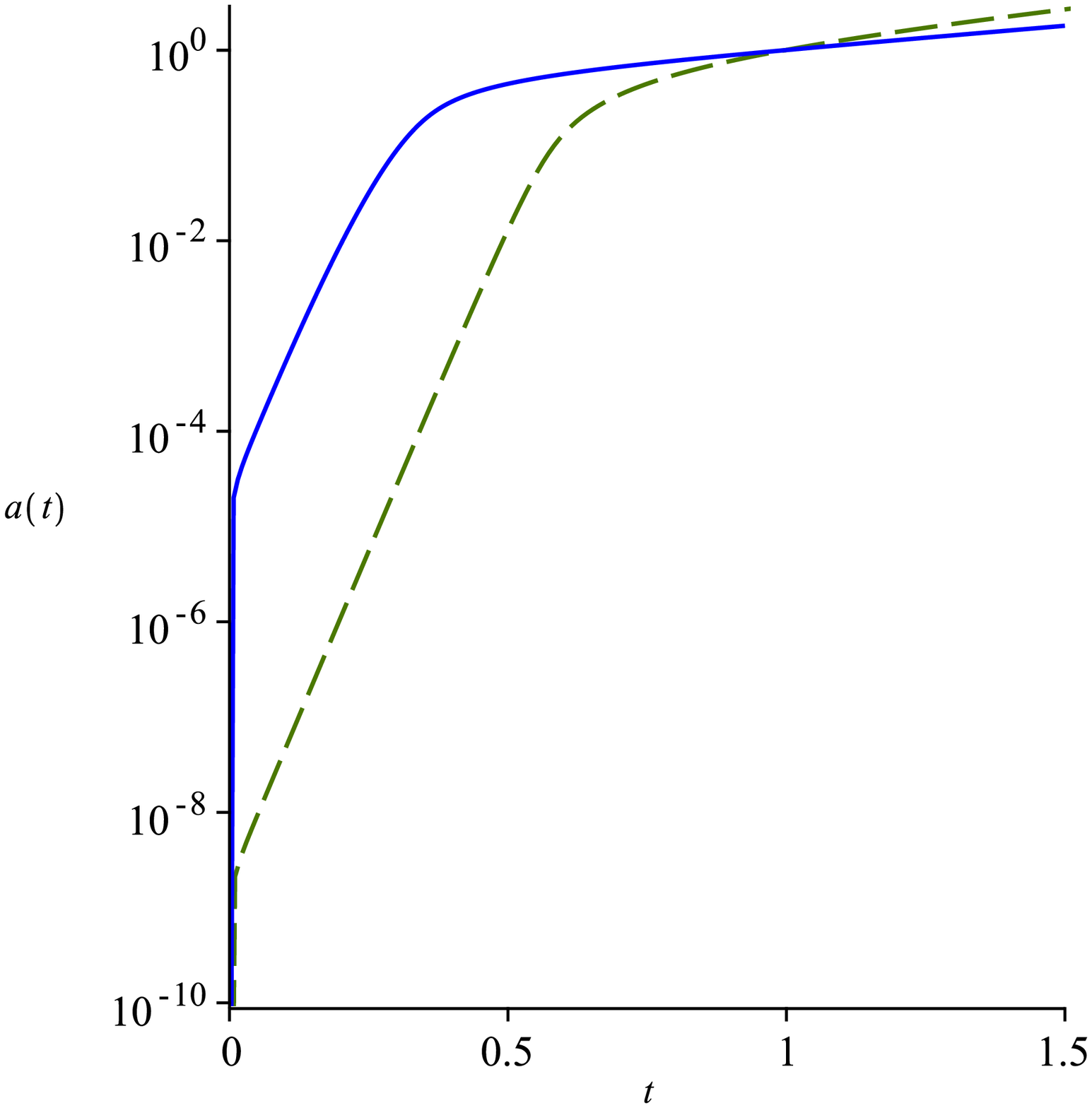} }}
       \subfloat{{\includegraphics[width=4cm]{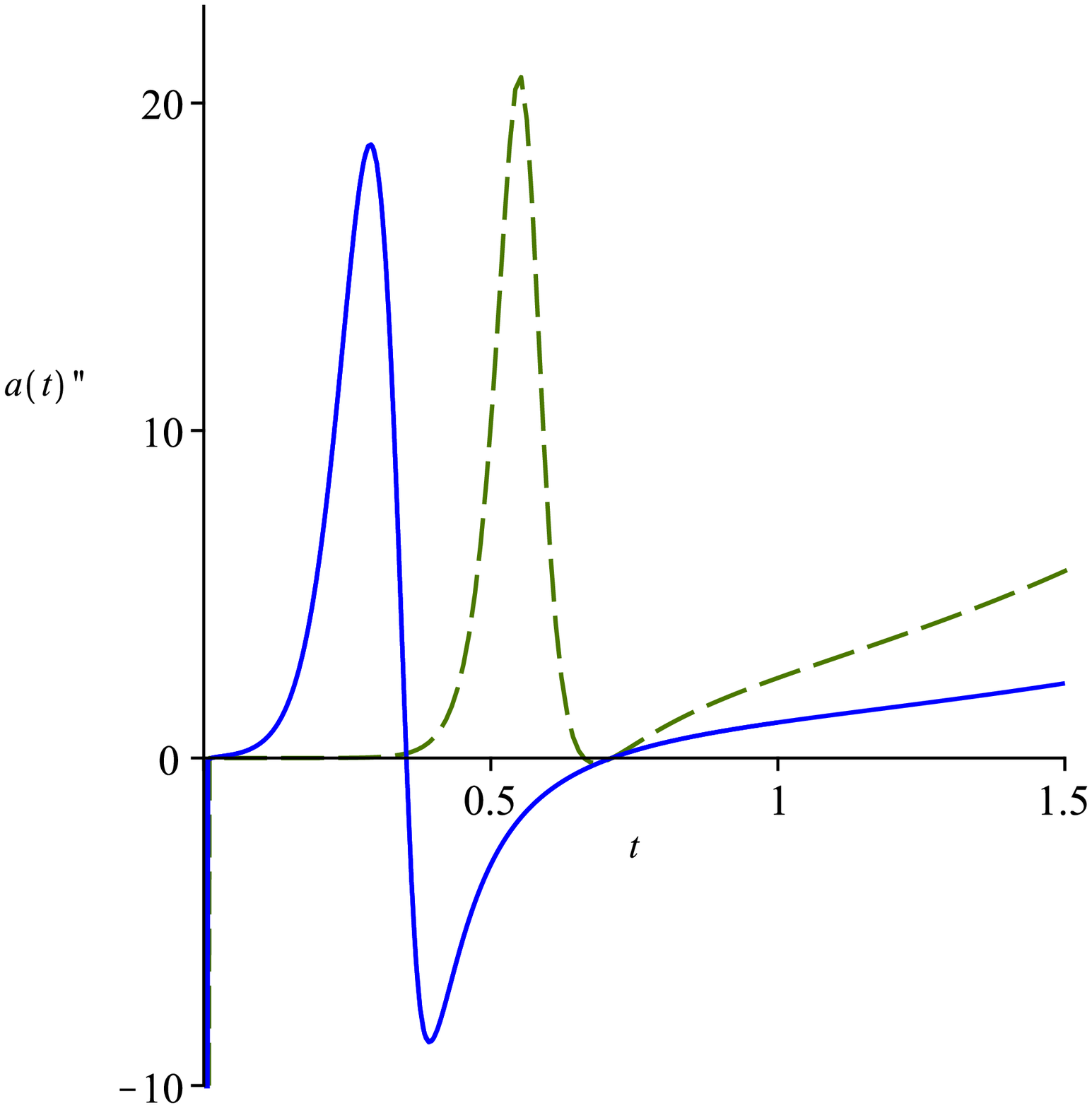} }}
        \subfloat{{\includegraphics[width=4cm]{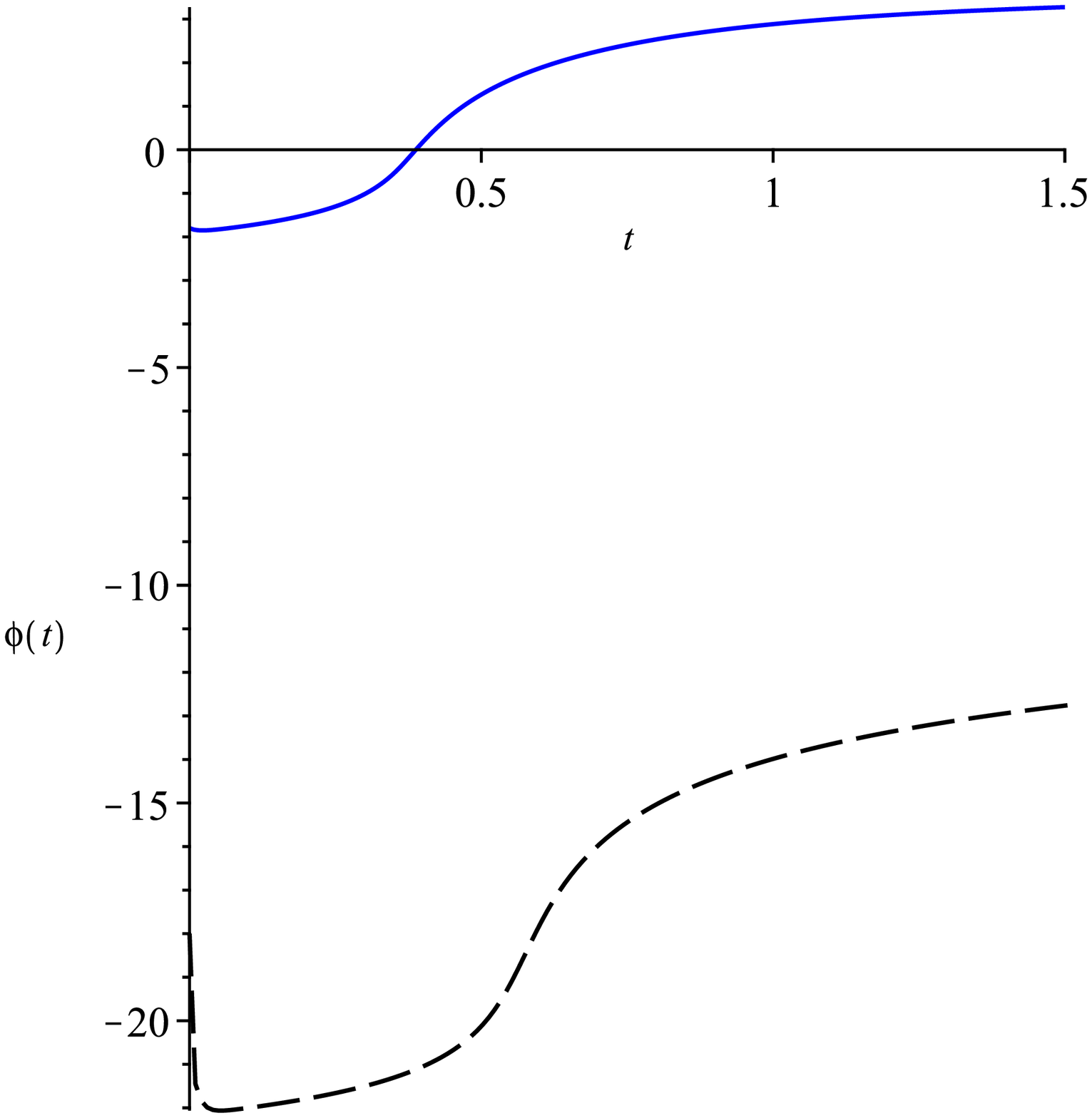} }}
       \\
      a\hspace{4cm} b \hspace{4cm} c\\
       \subfloat{{\includegraphics[width=4cm]{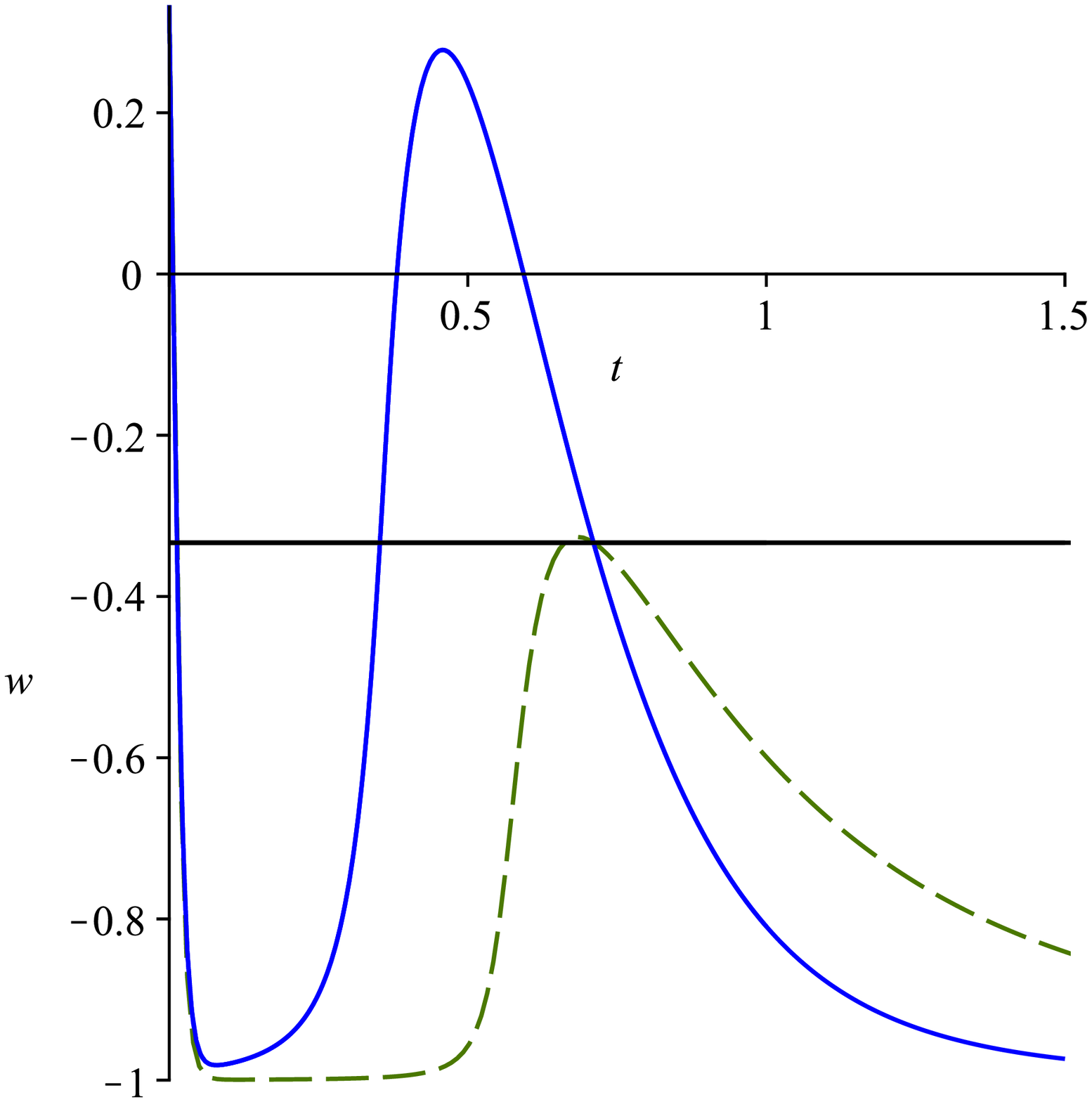} }}
        \subfloat{{\includegraphics[width=4cm]{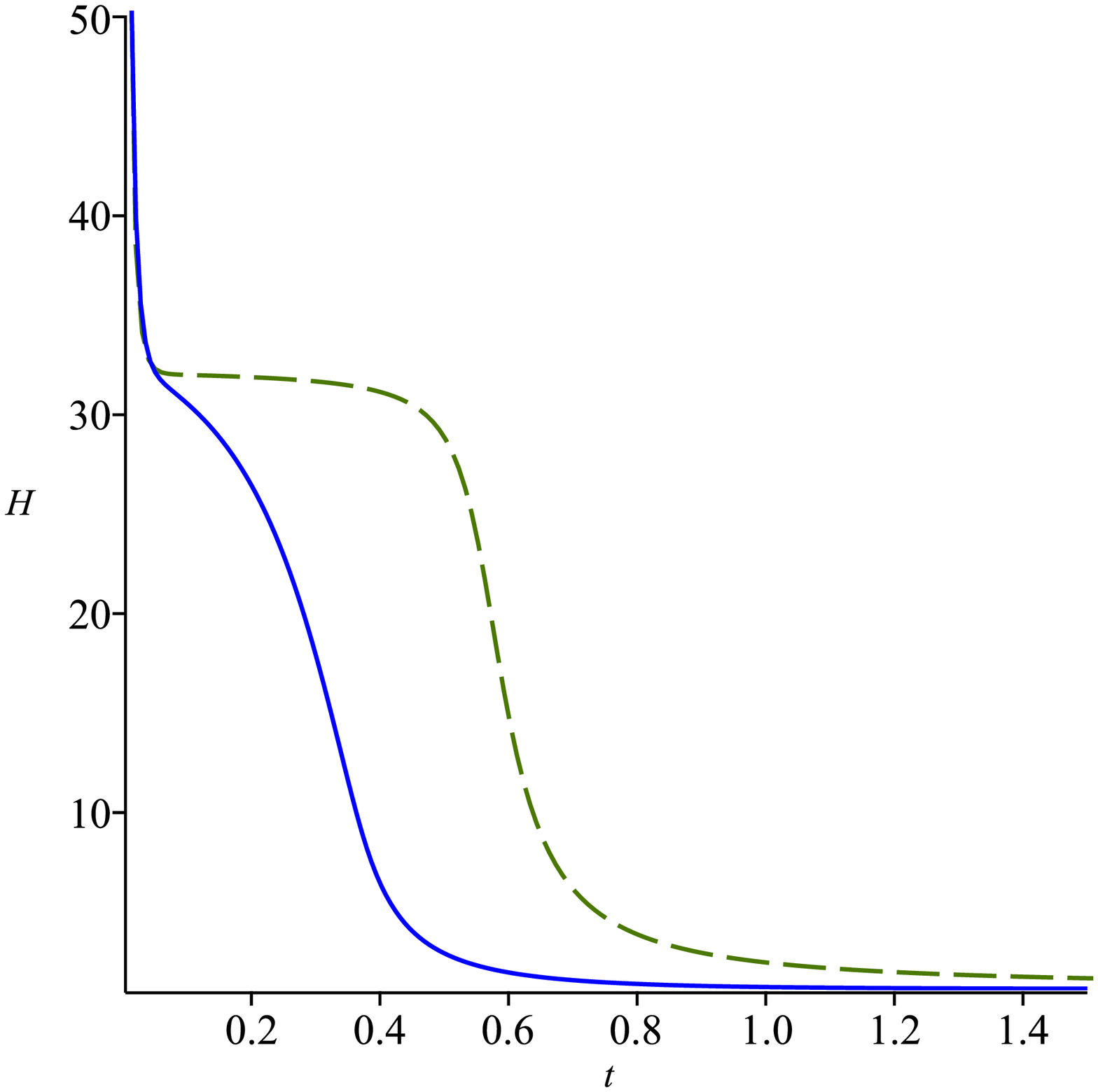} }}
      \subfloat{{\includegraphics[width=4cm]{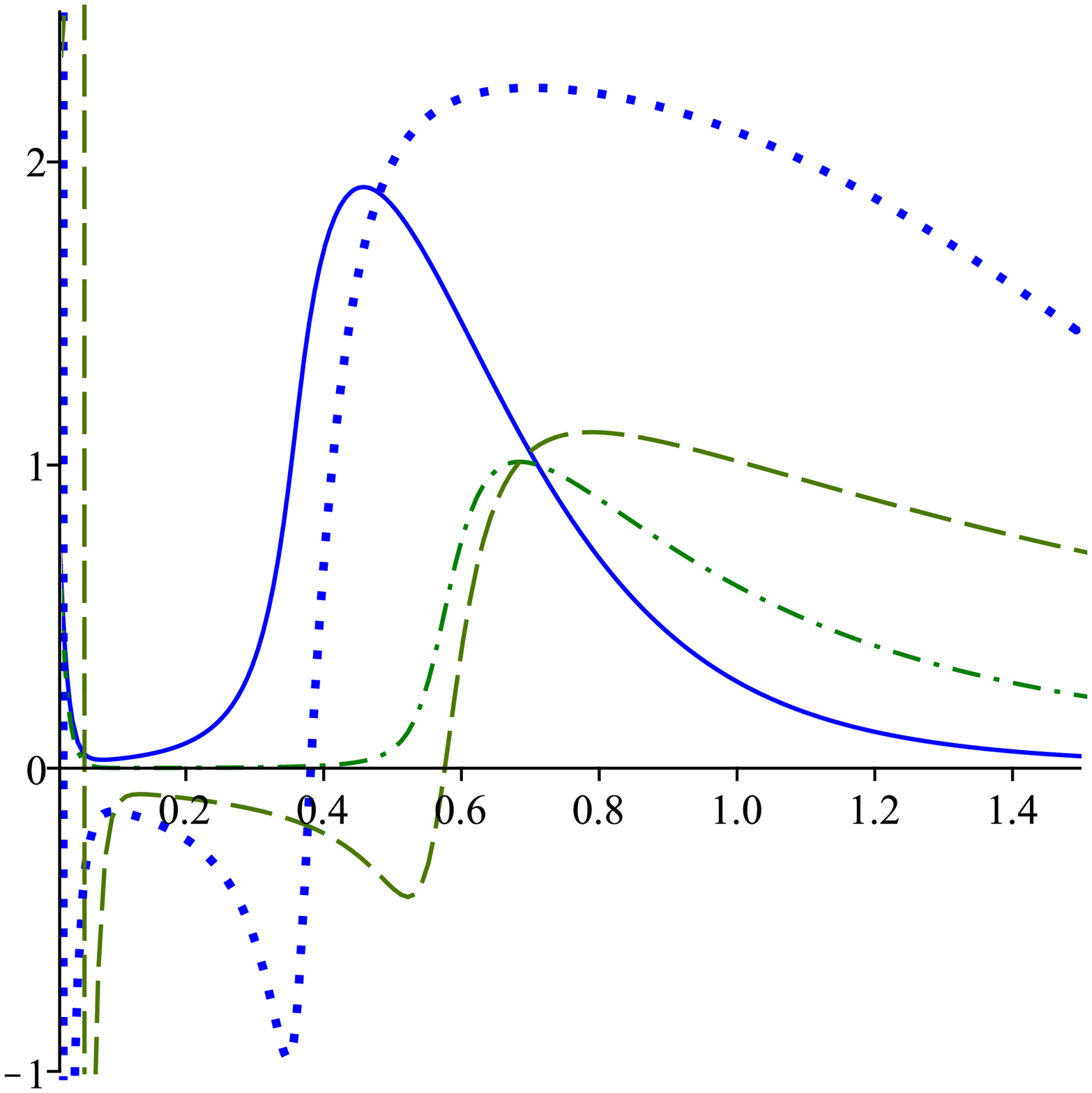} }}\\
      d \hspace{4cm}  e\hspace{4cm} f
       }
   \caption{
The Universe evolution: Case 1 (solid lines): $\chi_2=1,\;M_0=-0.04,\;M_1=1.53,\;M_2=10^{-3},\;\alpha=1.4,\;b_0=0.016,\;p_u=11.5\times 10^{-12},\; f_1=5.86,\; f_2=10^{-3},\;\phi(0)=-1.8$.
 Case 2 (dashed lines): $\chi_2=4\times10^{-5},\;M_0=-0.01,\;M_1=0.763,\;M_2=4,\;\alpha=0.64,\;b_0=1.52\times 10^{-7},\;p_u=6.5\times 10^{-24},\; f_1= 10^{-4},\; f_2=10^{-8},\;\phi(0)=-18$.  
 On the panels are the evolution of: 
 (a) the logarithm of the scale factor $a(t)$, 
 (b) the second derivative of scalar field  $\ddot{a}(t)$, 
 (c) the inflaton field $\phi(t)$, 
 (d) the equation of state $w(t)=p(t)/\rho(t)$, 
 (e) the Hubble parameter $H$
 and (f) the slow  roll parameter $\epsilon$ (Case 1: solid line, Case 2: dashes) and $\eta$ (Case 1: dots Case 2: dash-dotted line).
 }  
   \label{Fig1}
     \end{figure}

In order to illustrate  the physically realistic evolution we consider two different and hopefully representative regions of the parameter-space.
In the first case the dimensionless parameter $\chi_2 \sim 1$ ($M_2<<\vert M_0\vert$) and in the second one $\chi_2 << 1$ ($M_2\gg\vert M_0 \vert$).
The exact values of all other  parameters are given in the Fig.\ref{Fig1} caption.
We have used the parameter $b_0$ to set $t_{\mathrm{AE}}\sim0.71$ and parameters $f_1$ and $p_u$ to ensure $a(1)=1$.

The two different region of exponential expansion of the Universe are well visible on the logarithmic plot of $a(t)$ Fig \ref{Fig1} panel (a). 

The other phases of the Universe evolution are easily traceable on Fig. \ref{Fig1}, panel (b) where the time dependence of $\ddot{a}(t)$ is plotted, and also from panel (d) where the equation of state (EOS)  $w=p / \rho$ of the universe  is presented. 
 We can see that the Universe passes trough the following stages: 
First, beginning at $t=0$ we observe the EOS of ultra-relativistic matter with $w=1/3$.
The existence of this phase doesn't contradict to the observations, because currently we have information solely from the time after the initial inflation.
Second, after the moment $t_{EI}$ we have inflation with EOS of dark energy $w \to -1$.
Third, after the moment $t_{SD}$ the Universe is in matter domination stage where $w>-1/3$ and $w\to 0$.
Finally, after the moment $t_{AE}$  we have slow accelerated expansion with $w<-1/3$.
These numerical results coincide with the theoretically calculated asymptotic for $w$, namely $w\vert_{a(t)=0}\to 1/3$ and $w\vert_{a(t)=\infty}=-1$.

The friction in the model can be observed on Fig. \ref{Fig1}, panels (c), where it is seen that the inflaton scalar field $\phi$ tends to a constant.
This means that the theory predicts the existence of a scalar field with in general nonzero average value in the late Universe.

On the final two panels  (e) and (f) of Fig. \ref{Fig1} we present the logarithm of the Hubble parameter $H=\dot{a}(t)/a(t)$ and the ``slow-roll'' parameters of the model \cite{9408015} defined as  $\epsilon=-\dot{H}/H^2$ and $\eta=-\ddot{\phi}/H\dot{\phi}$. 
We observed that both $\eta$ and $\epsilon$ are $<|1|$ during inflation even though, they are not much smaller than $1$ as expected. 
The parameter $\eta$ is  $\eta\sim 0.1$  during the inflation  in both cases, but at the beginning of the  accelerated expansion  $\eta<1$ only in Case 2. 
We can attribute this to the fact that during the inflation in Case 1 we have $A(\phi)>>1$, while in Case 2 $A(\phi)\to 1$ (see Eq. \ref{eq_infl}).
In Case 2, the parameter $\epsilon_{EI} \to 0$  and $\epsilon_{AE}<0.5$, while in Case 1, $\epsilon_{EI}\to 0.1$ and $\epsilon_{AE}\sim 1$. Since $\epsilon$ measures the "exponentiality" of $a(t)$, this shows that in Case 2 the dynamics is dominated entirely by inflaton in nearly constant potential, while in Case 1 the dynamics of the inflaton and darkon are not decoupled.  
Physically, we attribute the observed difference mainly to the more complex movement of the inflaton in this case --- it initially moves to smaller values, then at $t\sim 0.05$ it turns back and gradually increases its value (see Fig. \ref{Fig1} (c)). Thus for relatively long period of time $U_{eff}(\phi(t))$ does not change significantly which together with the small value of $p_u$ ensures almost pure exponential growth of $a(t)$.

The parameter $\alpha$ is crucial for the length of both  inflation  and  matter domination epochs. 
We have investigated the model behavior with respect to small parameter variation as well.
 In both cases, higher value of $\alpha$ leads to a lower $t_{SD}$ and therefore longer matter domination. However, it turns out that it is impossible to obtain realistic $t_{\mathrm{SD}}$ solely by increasing $\alpha$. Two other parameters $f_1$ and $f_2$ affect significantly the value of $t_{SD}$.
 When they are increased we get smaller  $t_{SD}$. As a result, we expect $\alpha,\;\;f_1$ and $f_2$ to have really big values for  realistic $t_{SD}$. This is just the opposite to the estimates in Ref.\cite{1408.5344} where it is asserted that in order for the theory to be compatible with Planck data \cite{cosmo6}, one needs $\alpha\to 0$ and in addition $f_1\sim f_2 \ll 1$. From  our experiments shown on the plots, the smaller $\alpha$ (Case 2) leads to significantly longer early inflation, thus leading to very short matter-domination epoch. 

The e-folds parameter related to the inflation is defined as $N=\ln{(a_{SD}/a_{EI})}$ where the subscripts denotes the beginning ($(EI)$) and end ($(SD)$) of the inflation. 
The theoretical estimation for the number of e-folds needed to solve the horizon problem is model-dependent but is $N>70$. One such estimation can be found in Ref. \cite{heeck2011} .
 In our cases, $t_{EI}=0.013, t_{SD}= 0.349$ (Case 1) and $t_{EI}=0.014, t_{SD}=0.663$ (Case 2),  and thus 
$N^{Case 1}\approx 9$ and $N^{Case2}\approx 18$. The higher e-folds number in Case 2 is easily explained by the much longer inflation in this case. For the Hubble parameter, we obtain at $t=1$, in the two cases $H_0=\{1.2,2.4\}$ respectively. One can see that in the first case, the modern-day Hubble parameter is much closer to the theoretically expected value, due to the fact that in the first case, the inflaton is lower on the right plateau of the effective potential, thus $U_{eff}\to U_+$. In the second case, we are a bit higher on the potential, which affects the value of $H$. From the numerical experiments it seems that obtaining better Hubble and e-folds parameters is a matter of extensive numerical study of the full parameter space, which we have not performed yet. 

\section{Conclusions}
We have presented the numerical study of the model of Guendelman-Nissimov-Pacheva aiming to describe inflation, dark energy, dark matter in a seamless way trough the evolution of two scalar fields -- the darkon and the inflaton. 

Our calculations have confirmed that qualitatively this model can successfully describe the evolution of Universe, since it naturally reproduces its main phases. The values of the parameters we use, however, differ significantly from the ones estimated in Refs. \cite{1408.5344, 1609.06915}, which poses important questions related to the nature of those parameters. 

Furthermore, we have shown that there is a strong friction term in the system, due to which, the inflaton field tends to a constant. This means that the model predicts the existence of a non-zero averaged scalar field in the current Universe. 

Also, we have shown that the inflationary region of the effective potential is its slope, while the evolution starting from higher plateau is nonphysical as it leads to eternal inflation. As a result, the duration of the inflation in the ``realistic'' Universe evolution is always finite, contrary to the case in Refs.\cite{1408.5344, 1609.06915} where the inflation can be arbitrary long.

Finally, an interesting feature is that for certain choices of the parameters and the initial conditions of the system, the inflaton field increases its value in the beginning of the integration. This hints of a dynamical mechanism for enhancing the initial exponential expansion by holding the inflaton near its starting position on the effective potential. 

\section*{Acknowledgments}
It is a pleasure to thank E. Nissimov and S. Pacheva for the discussions. The work is supported by BAS contract DFNP -- 49/21.04.2016, by Bulgarian NSF grant DN-18/1/10.12.2017 and by Bulgarian NSF grant 8-17.



\end{document}